\documentclass[pra,twocolumn,longbibliography]{revtex4-2}
\usepackage{amsmath}
\usepackage{amssymb}
\usepackage{graphicx} 

\usepackage[usenames, dvipsnames]{color}
\usepackage[colorlinks=true,citecolor=blue,urlcolor  = purple]{hyperref}

\newcommand{\ket}[1]{\vert #1 \rangle}

\newcommand{\ketbra}[2]{\vert #1 \rangle \langle #2 \vert}
\newcommand{\abs}[1]{\vert #1 \vert}
\newcommand{\tr}{\operatorname{tr}}
\newcommand{\dprod}[2]{\langle #1, #2 \rangle}

\newcommand{\mean}[1]{\langle #1 \rangle}

\newcommand{\RR}{\mathbb{R}}
\newcommand{\CC}{\mathbb{C}}

\usepackage{xcolor}

\begin{document}
\title{Shadow tomography with noisy readouts}
\author{Hai-Chau Nguyen}
\email{chau.nguyen@uni-siegen.de}

\affiliation{Naturwissenschaftlich--Technische Fakult\"{a}t,
Universit\"{a}t Siegen, \\ Walter-Flex-Stra{\ss}e 3, 57068 Siegen, Germany}
\date{\today}

\begin{abstract}
Shadow tomography is a scalable technique to characterise the quantum state of a quantum computer or quantum simulator. 
The protocol is based on the transformation of the outcomes of random measurements into the so-called classical shadows, which can later be transformed into samples of expectation values of the observables of interest. 
By construction, classical shadows are intrinsically sensitive to readout noise. 
In fact, the complicated structure of the readout noise due to crosstalk appears to be detrimental to its scalability. 
We show that classical shadows accept much more flexible constructions beyond the standard ones, which can eventually be made more conformable with readout noise.   With this construction, we show that readout errors in classical shadows can be efficiently mitigated by randomly flipping the qubit before, and the classical outcome bit after the measurement, referred to as $X$-twirling. 
That a single $X$-gate is sufficient for mitigating readout noise for classical shadows is in contrast to Clifford-twirling, where the implementation of random Clifford gates is required.
\end{abstract}

\maketitle

The success in fabrication of quantum computers and quantum simulators with increasing number of qubits in the last few years has put a strong demand on development of methods to characterise their output quantum state~\cite{arute_quantum_2019,Kim2023a}.
Conventional state tomography fails utterly due to its exponential complexity as the system size is large~\cite{smithey_wigner_dist,james_measurement_qubit,heaffner_multiparticle_ions,schwemmer_errors_tomography,paris_quantum_state_estimation,guta_fast_2020} and 
shadow tomography has been proposed as a scalable alternative~\cite{aaronson_journal_2020,huang_predicting_2020}.  
The basic idea of shadow tomography is to replace the traditional description of the simulated quantum state by measuring its exponentially large density operator with sampling the so-called classical shadows. 

\begin{figure}[t]
    \centering
    \includegraphics[width=0.45\textwidth]{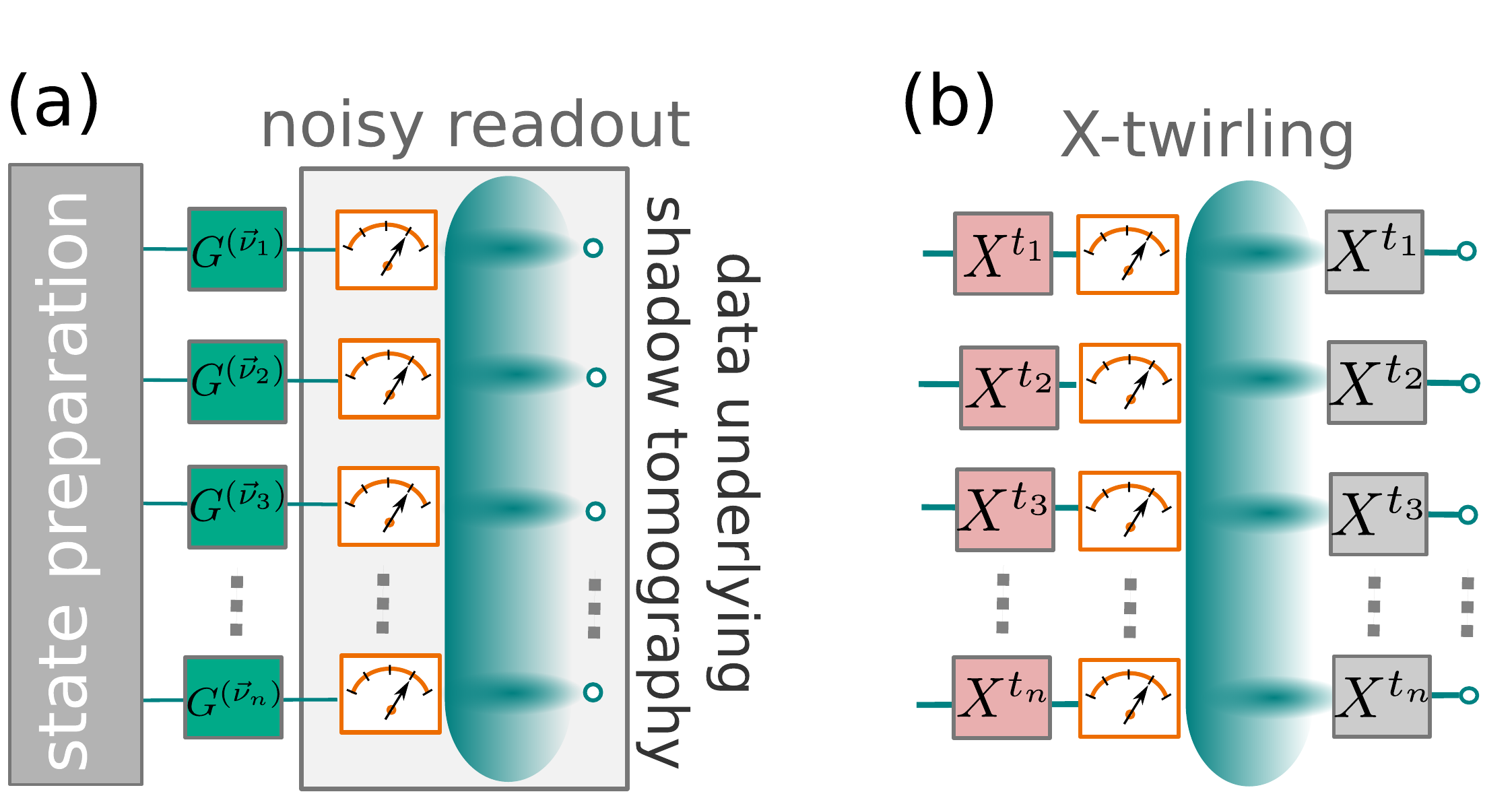}
    \caption{(a) \textbf{Shadow tomography with noisy readout including crosstalk.} State prepared by the quantum simulator is subsequently subject to random unitaries $G^{(\vec{\nu}_i)}$ so that the final measurement in the computational basis corresponds to measurement in a random direction $\vec{\nu}_i$. Readout noise eventually introduces correlated bit-flips to the collected data for shadow tomography. (b) \textbf{$X$-twirling of the readout noise.} Randomly (if $t_i=1$), a quantum $X$ gate is applied to qubit $i$  right before the measurement; the obtained classical bit is flipped by the classical $X$ gates accordingly. } 
    \label{fig:noisy-st}
\end{figure}

In shadow tomography,  the state of the $n$ qubits of the quantum simulator is sampled by measuring in 
a random basis indexed by $\pmb{\nu}$ among $N$ possible choices. 
In practice, a particular basis $\pmb{\nu}$ is implemented by subjecting the qubits to an appropriate evolution followed by a measurement in the computation basis, yielding an outcome bitstring $\pmb{s}$. 
The pair $(\pmb{\nu},\pmb{s})$ for each run is then referred to as a \emph{generalised outcome} and $\ketbra{\pmb{\nu},\pmb{s}}{\pmb{\nu},\pmb{s}}$ denotes the projection onto outcome $\pmb{s}$ in basis $\pmb{\nu}$.

Notice that for a given quantum state $\rho$, the generalised outcome $(\pmb{\nu},\pmb{s})$ can be considered as a random variable distributed according to the distribution $p(\pmb{\nu},\pmb{s}) = \tr(\rho P^{\pmb{\nu}}_{\pmb{s}})$ where $P^{\pmb{\nu}}_{\pmb{s}} =1/N \ketbra{\pmb{\nu},\pmb{s}}{\pmb{\nu},\pmb{s}}$.
In shadow tomography, each generalised outcome $(\pmb{\nu},\pmb{s})$ is associated with a so-called \emph{classical shadow} $\rho^{\pmb{\nu}}_{\pmb{s}}$ in the state space satisfying the unbiasedness  condition
\begin{equation}
    \sum_{\pmb{\nu},\pmb{s}} p(\pmb{\nu},\pmb{s}) \rho^{\pmb{\nu}}_{\pmb{s}} = \rho.
    \label{eq:fundamental}
\end{equation}
Sampling the random generalised outcome  $(\pmb{\nu},\pmb{s})$ thus leads to sampling the random classical shadow $\rho^{\pmb{\nu}}_{\pmb{s}}$, the mean of which converges to $\rho$. 

Notice that the choice of the measurements put constraints on the possible classical shadows by equation~\eqref{eq:fundamental}. 
Crucially, in shadow tomography the measurements have to be designed such that the classical shadows can be efficiently presented. 
The single qubit randomised measurement design is arguably the most practical scheme for the current and near future technology due to its simplicity~\cite{huang_predicting_2020}.
In this scheme, the random measurement $\pmb{\nu}$ is implemented by measuring each qubit separately in a random direction $\vec{\nu}_i$ uniformly sampled from a certain predefined set $\mathcal{S}$. 
Practically, measurement of qubit $i$ in direction $\vec{\nu}_i$ is implemented by acting an appropriate local unitary gate $G^{(\vec{\nu}_i)}$ before the measurement in the computational basis; see Fig.~\ref{fig:noisy-st}a. 
The standard construction~\cite{huang_predicting_2020,Nguyen2022a} then leads to classical shadows factorising over the qubits, generically written as
\begin{equation}
    \rho^{\pmb{\nu}}_{\pmb{s}} = \otimes_{i=1}^{n} \frac{1}{2}[\openone + (-1)^{s_i} \xi^{\vec{\nu}_i}],
    \label{eq:unmitigated-shadow}
\end{equation}
for certain traceless operator $\xi^{\vec{\nu}_i}$, which can be explicitly computed given the directions $\mathcal{S}$~\cite{Nguyen2022a}; see Appendix~A. 
For example, $\mathcal{S}=\{x,y,z\}$ corresponds to the Pauli measurements and $\xi^{\alpha} = 3 \sigma^{\alpha}$, where $\sigma^{\alpha}$ with $\alpha=x,y,z$ are the Pauli observables.
Notice that the single qubit classical shadows in the tensor factor of~\eqref{eq:unmitigated-shade} reflects the inherent inversion symmetry of two projections in a single measurement setting~\cite{nguyen_symmetries_2020}.

These classical shadows can be used to estimate correlation functions from the simulated quantum state. 
Such a correlation function is the expectation value of an observable acting over a group of few qubits, hereafter called correlator. 
A correlator can be specified by a pattern bitstring $\pmb{v}= \{v_i\}$ and a collection of directions $\pmb{\mu}=\{\vec{\mu}_i\}$, where $v_i=1$ indicates its non-trivial action on qubit $i$ with observables $\sigma^{\vec{\mu}_i}= \vec{\mu}_i \cdot \vec{\sigma}$, where $\vec{\sigma} = (\sigma^x,\sigma^y,\sigma^z)$.  
Explicitly, such a correlator
$C^{\pmb{\mu}}_{\pmb{v}}$ can be written as
\begin{equation}
    C^{\pmb{\mu}}_{\pmb{v}} = \otimes_{i=1}^{n} [\delta_{v_i,0} \openone + \delta_{v_i,1} \sigma^{\vec{\mu}_i}].
    \label{eq:correlator}
\end{equation}

The classical shadow $\rho^{\pmb{\nu}}_{\pmb{s}}$ as random variable defines a random variable $c^{\pmb{\mu},\pmb{\nu}}_{\pmb{v},\pmb{s}} = \tr [\rho^{\pmb{\nu}}_{\pmb{s}} C^{\pmb{\mu}}_{\pmb{v}}]$ for the correlator $C^{\pmb{\mu}}_{\pmb{v}}$, which we might call a \emph{shade}. 
Thus sampling the random classical shadows can eventually be converted to sampling shades for a  correlation function. 
That the mean of the random shade $\mean{c^{\pmb{\mu},\pmb{\nu}}_{\pmb{v},\pmb{s}}}$ corresponds to the correlation function $ \mean{C^{\pmb{\mu}}_{\pmb{v}}}$ follows directly from the unbiasedness  condition for classical shadows~\eqref{eq:fundamental}. Crucially to the scalability of shadow tomography, the shades provided by the classical shadows~\eqref{eq:unmitigated-shadow} 
 for the correlator~\eqref{eq:correlator} can be efficiently computed,
\begin{equation}
    c^{\pmb{\mu},\pmb{\nu}}_{\pmb{v},\pmb{s}} =  \prod_{i=1}^{n}[\delta_{v_i,0} + (-1)^{s_i} \delta_{v_i,1} \frac{1}{2} \tr (\xi^{\vec{\nu}_i} \sigma^{\vec{\mu}_i}) ].
    \label{eq:unmitigated-shade}
\end{equation}
In this way, correlation functions can be estimated without ever constructing the exponentially large density operator explicitly~\cite{huang_predicting_2020}.

Since its discovery, applications of shadow tomography have been developed in a very fast pace. 
Small sampling of such applications includes energy estimation~\cite{hadfield_adaptive_2021,hadfield_measurements_2020}, entanglement detection~\cite{elben_mixed-state_2020,neven_symmetry-resolved_2021}, metrology~\cite{rath_quantum_2021},  analysing scrambled data~\cite{garcia_quantum_2021} and quantum chaos~\cite{joshi_probing_2022}. 
Improvements of the performance and scope of the scheme have also been made in parallel~\cite{huang_efficient_2021,elben_mixed-state_2020,zhang_experimental_measurement_shadows_2021,chen_robust_shadow_estimation_2021,hu_hamiltonian_driven_2022,hu_scrambled_dynamics_2022,levy_classical_2021,helsen_estimating_2021}.
Despite this significant progress, one fundamental problem of shadow tomography remains rather less understood.
The original shadow tomography protocol~\cite{huang_predicting_2020} assumed perfect operational quantum devices. 
Unfortunately, for the near-term quantum computers and simulators, neither quantum gates nor measurements are perfect~\cite{arute_quantum_2019,Chen2019}.
In fact, it is generally expected that mitigation of errors is crucial to practical applications of any quantum information processing protocol using the current and near-future quantum technology~\cite{arute_quantum_2019,Kim2023a,kandala2019error,kim2023scalable,o2022purification,Huang2022a,Cai2022a}; shadow tomography is of no exception~\cite{Jnane2023a,chen_robust_shadow_estimation_2021,Nguyen2022a,Koh2022a}.

On the one hand, gate imperfection strongly affects the quality of the target quantum state prepared by the simulator. 
In this aspect, it has been shown that the mitigation of gate errors in preparation of the target state by means of popular mitigation techniques for state preparation can be naturally followed by shadow tomography without much changes in principle~\cite{Jnane2023a}.
On the other hand, readout errors directly influence the shadow tomography as measurements are intrinsic to the data underlying classical shadows; see Fig.~\ref{fig:noisy-st}a.
A practical protocol for shadow tomography is thus expected to include the mitigation of readout errors intrinsically.

Several attempts in addressing readout errors for shadow tomography have been made in the last few years~\cite{Nguyen2022a,Koh2022a}.
If the errors are assumed to happen independently at each qubit at readout, they can also be efficiently characterised~\cite{Chen2019}. 
With this independent-flips model, it is shown that the errors can be mitigated without significantly altering the shadow tomography protocol and its scalability~\cite{Nguyen2022a}. 
However, it has recently become clear that this independent-flips model is too restrictive to fully capture the readout noises in the available quantum devices; 
crosstalk of readout errors generally need to be addressed~\cite{Bravyi2021a,Berg2022a}. 
Unfortunately, the crosstalk in readout errors is hard to characterise~\cite{Chen2019,Bravyi2021a}.
Moreover, it is also clear that crosstalk in readouts directly breaks the tensor factorising structure of conventional classical shadows~\eqref{eq:unmitigated-shade}, thus detrimental to its scalability; see again Fig.~\ref{fig:noisy-st}a. 

Recently, a significant step in resolving the readout errors for shadow tomography has been made~\cite{chen_robust_shadow_estimation_2021}. It is realised that if the measurement settings of shadow tomography are made by implementing the random unitaries sampled from an appropriate group, e.g., the single qubit Clifford group, the readout noise is effectively `twirled' to carry sufficient symmetry. As a result, the twirled noise can be in principle efficiently characterised and mitigated~\cite{chen_robust_shadow_estimation_2021}.  
However, repeatedly sampling random unitaries even from a simple group such as the single qubit Clifford group still poses a significant challenge for experiments. 

We are to show that as long as only readout errors are concerned, twirling by applying a single $X$-gate  randomly on the qubits as illustrated in Fig.~\ref{fig:noisy-st}b is in fact sufficient. 
This protocol is referred to as $X$-twirling. 
This is inspired by another recent breakthrough in realising that $X$-twirling is sufficient to mitigate readout errors for direct measurements of observables~\cite{Berg2022a}. 
That shadow tomography can be mitigated by $X$-twirling remains however non-trivial, as the twirled noise still contains crosstalk which breaks the tensor product structure~\eqref{eq:unmitigated-shadow} of classical shadows.
To overcome this difficulty, making uses of the flexibility in defining classical shadows by equation~\eqref{eq:fundamental}, we leave the standard constructions of classical shadows~\cite{huang_predicting_2020,Nguyen2022a,Koh2022a} and adopt an appropriate definition for mitigated classical shadows. 
Further, we show that the resulted classical shadows admit an efficient presentation in  the Fourier space, which eventually allows for scalable mitigation.

\textit{Readout-mitigated classical shadows---} Readout errors are typically modelled by a two-step process.
In the actual measurement, it is supposed that the ideal, exact outcome $\pmb{s}'$ of the measurement is generated, but cannot be observed.
The actual observed outcome $\pmb{s}$ is obtained by altering $\pmb{s}'$ with certain transition probability $R(\pmb{s} \vert \pmb{s}')$, which depends on the details of the device.
Thus if $p(\pmb{\nu},\pmb{s}')$ denotes the distribution of the ideal unobserved generalised outcome $(\pmb{\nu},\pmb{s}')$, the distribution of the actual observed outcome is $q(\pmb{\nu},\pmb{s})= \sum_{\pmb{s}'} R(\pmb{s} \vert \pmb{s}') p(\pmb{\nu},\pmb{s}')$. 
While this model may not capture all the physics of readout errors, it proves to be useful and has become the arguably most widely used model for readout errors for superconducting quantum computers~\cite{Chen2019}.

The standard constructions of classical shadows~\cite{huang_predicting_2020,Nguyen2022a,Koh2022a} unfortunately do not allow for a scalable mitigation of readout errors with crosstalk. 
However the definition of classical shadows satisfying~\eqref{eq:fundamental} is rather flexible.
We suggest the following definition of readout-mitigated classical shadows 
\begin{equation}
    \tilde{\rho}^{\pmb{\nu}}_{\pmb{s}} =  \sum_{\pmb{s}'} \rho^{\pmb{\nu}}_{\pmb{s}'} R^{-1} (\pmb{s}' |\pmb{s}),
    \label{eq:mitigated-shadow}
\end{equation}
where $\rho^{\pmb{\nu}}_{\pmb{s}'}$ are the unmitigated classical shadows in equation~\eqref{eq:unmitigated-shadow} and $R^{-1}$ denotes the inverse of the transition probabilities as a matrix. 
The inverse of the transition matrix in definition~\eqref{eq:unmitigated-shadow} is deliberately used to naturally cancel the readout noise so that the unbiasedness condition~\eqref{eq:fundamental} remains valid for the readout-mitigated classical shadows, $\sum_{\pmb{\nu},\pmb{s}} q(\pmb{\nu},\pmb{s}) \tilde{\rho}^{\pmb{\nu}}_{\pmb{s}} = \rho$; see Appendix~B.

Definition~\eqref{eq:unmitigated-shadow} remains however formal, as the transition matrix itself $R(\pmb{s}' |\pmb{s})$ cannot be measured due to its exponential size, let alone its inversion.
To overcome this issue, we make use of the idea of $X$-twirling~\cite{Berg2022a}. 
Here a random bitstring $\pmb{t} \in \{0,1\}^{n}$ is drawn. 
An $X$-gate is applied to the qubit before the measurement if $t_i=1$, and the obtained outcome classical bit is also flipped after the measurement; see Fig.~\ref{fig:noisy-st}b. 
In this way, the readout noise in fact acts on a random bitstring. 
Mathematically, this leads to the replacement of the transition matrix $R(\pmb{s}|\pmb{s}')$ by its average over random bitstrings, $\bar{R}(\pmb{s} | \pmb{s}')=1/2^n \sum_{\pmb{t}}R(\pmb{s} \oplus \pmb{t}|\pmb{s}' \oplus \pmb{t})$, where $\oplus$ denotes the addition modulo $2$; the detailed discussion is given in Appendix~C.

Notice that the space of bitstrings $\{0,1\}^n$ can be considered as a vector space of dimension $n$ over the field $\{0,1\}$. A bitstring $\pmb{t} \in \{0,1\}^n$ in fact defines a translation in this space, mapping a bitstring $\pmb{s}$ to $\pmb{s} \oplus \pmb{t}$.
The twirled transition matrix $\bar{R}(\pmb{s} | \pmb{s}')$ has an important property that it is translationally invariant, $\bar{R}(\pmb{s} | \pmb{s}') = \bar{R}(\pmb{s} \oplus \pmb{t} | \pmb{s}' \oplus \pmb{t})$. 

Further, the space $\{0,1\}^n$ also has a Fourier kernel given by $(-1)^{\dprod{\pmb{w}}{\pmb{s}}}$ with the identity $\sum_{\pmb{s}} (-1)^{\dprod{\pmb{w}}{\pmb{s}}} = 2^n \delta_{\pmb{w},\pmb{0}}$, where the scalar product is defined as usual, $\dprod{\pmb{w}}{\pmb{s}} = \sum_{i=1}^{n} w_i s_i$.
While this Fourier kernel has somewhat peculiar form, it mimics the familiar Fourier transform  in physics in many aspects.
In particular, as the transition matrix $\bar{R}(\pmb{s} | \pmb{s}')$ is translational invariant, it can be expected to be \emph{diagonal} in the Fourier space. 
Indeed, one can write
\begin{equation}
    \bar{R} (\pmb{s}|\pmb{s}') = \frac{1}{2^n}\sum_{\pmb{w}} (-1)^{\dprod{\pmb{w}}{\pmb{s} \oplus \pmb{s}'}} g(\pmb{w}),
    \label{eq:fourier-noise}
\end{equation}
where $g(\pmb{w}) = \sum_{\pmb{s}} (-1)^{\dprod{\pmb{w}}{\pmb{s}}} \bar{R} (\pmb{s} | \pmb{0})$. As such, the transition matrix $\bar{R} (\pmb{s}|\pmb{s}')$ is completely characterised by its Fourier components $g(\pmb{w})$.

That the transition matrix admits an efficient representation in the Fourier space inspires that the mitigated classical shadows can also be efficiently presented in the Fourier space. 
To this end, we define the Fourier transform for the mitigated classical shadows, $\tilde{\tau}^{\pmb{\nu}}_{\pmb{w}}  = \sum_{s} (-1)^{\dprod{\pmb{w}}{\pmb{s}}} {\tilde{\rho}}^{\pmb{\nu}}_{\pmb{s}}$,
and for unmitigated classical shadows,
${\tau}^{\pmb{\nu}}_{\pmb{w}}  = \sum_{s} (-1)^{\dprod{\pmb{w}}{\pmb{s}}} {\rho}^{\pmb{\nu}}_{\pmb{s}}$, respectively.
Relation~\eqref{eq:mitigated-shadow} indeed becomes simple in the Fourier space,
\begin{equation}
    \tilde{\tau}^{\pmb{\nu}}_{\pmb{w}} = \frac{1}{g(\pmb{w})} \tau^{\pmb{\nu}}_{\pmb{w}}.
    \label{eq:fourier-mitigation}
\end{equation}
Using the unmitigated classical shadows $\rho^{\pmb{\nu}}_{\pmb{s}}$ in equation~\eqref{eq:unmitigated-shadow}, one can compute its Fourier transform $\tau^{\pmb{\nu}}_{\pmb{w}}$ explicitly. 
Equation~\eqref{eq:fourier-mitigation} then allows for computing the Fourier transform of the mitigated classical shadows $\tilde{\tau}^{\pmb{\nu}}_{\pmb{w}}$. 

For a correlator $C^{\pmb{\mu}}_{\pmb{v}}$, the computation of the mitigated shade $\tilde{c}^{\pmb{\mu},\pmb{\nu}}_{\pmb{v},\pmb{s}} = \tr (\tilde{\rho}^{\pmb{\nu}}_{\pmb{s}} C^{\pmb{\mu}}_{\pmb{v}})$ however requires the classical shadows in real space $\tilde{\rho}^{\pmb{\nu}}_{\pmb{s}}$. 
As such, it is an exponential summation over all wavevectors of the classical shadows in the Fourier space, $\tilde{c}^{\pmb{\mu},\pmb{\nu}}_{\pmb{v},\pmb{s}} = 1/2^{n} \sum_{\pmb{w}} (-1)^{\dprod{w}{s}} \tr ( \tilde{\tau}^{\pmb{\nu}}_{\pmb{w}} C^{\pmb{\mu}}_{\pmb{v}})$.
Fortunately, for the correlator of the form~\eqref{eq:correlator}, only the single classical shadow $\tilde{\tau}^{\pmb{\nu}}_{\pmb{w}}$ that has the wavevector $\pmb{w}$ matching the pattern $\pmb{v}$ contributes to this sum; see Appendix~D. We eventually obtain a simple formula for the shade
\begin{equation}
\tilde{c}^{\pmb{\mu},\pmb{\nu}}_{\pmb{v},\pmb{s}} =  \frac{(-1)^{\dprod{\pmb{v}}{\pmb{s}}}}{g(\pmb{v})} \prod_{i=1}^{n}  [\delta_{v_i,0}  +  \delta_{v_i,1}\frac{1}{2} \tr(\xi^{\vec{\nu}_i} \sigma^{\vec{\mu}_i})].
    \label{eq:mitigated-shade}
\end{equation}
That the mitigated  shades $\tilde{c}^{\pmb{\mu},\pmb{\nu}}_{\pmb{v},\pmb{s}}$ for a correlator  can be efficiently computed by means of~\eqref{eq:mitigated-shade} restores the scalability of shadow tomography under readout noise even in the present of crosstalk.
Notice that~\eqref{eq:mitigated-shade} is strikingly similar to formula~\eqref{eq:unmitigated-shade}. The factor $1/g(\pmb{v})$ also resembles the mitigation of observables by $X$-twirling~\cite{Berg2022a}.
It is however to be emphasised that these similarities are a non-trivial consequence of the symmetry of the unmitigated classical shadows~\eqref{eq:unmitigated-shadow}, the particular definition of the noisy classical shadows~\eqref{eq:mitigated-shadow}, the factorisation of the twirled noise in Fourier space~\eqref{eq:fourier-mitigation} and the matching condition underlying~\eqref{eq:mitigated-shade} for correlators of~\eqref{eq:correlator}.

\textit{Protocols and their complexity---} Our theoretical consideration above leads to the following protocol for $X$-twirled calibration aiming at estimating $g(\pmb{w})$, which in fact coincides with that for mitigation of observables~\cite{Berg2022a}:

\medskip
\begin{tabular}{l}
\textbf{$X$-twirled calibration} \\
\hline
\hline
Repeat the following steps $M_{\text{c}}$ times: \\ 

1. Prepare the system in $\ket{\pmb{0}}$ state \\

2. Draw a random bitstring $\pmb{t}=\{t_i\}$ \\

3. Apply an $X$ gate for every bit $i$ with $t_i=1$ \\

4. Measure the system in the computational basis \\

5. Flip the obtained bitstring at position $i$ with $t_i=1$ \\

6. Record the final bitstring $\pmb{s}$ \\
\hline
\hline
\textbf{Output:} Calibration bitstrings $\{\pmb{s}^{(k)}\}_{k=1}^{M_{\text{c}}}$.
\end{tabular}
The Fourier component  $g(\pmb{w})$ is then estimated by replacing the average over distribution $\bar{R} (\pmb{s} | \pmb{0})$ in its definition~\eqref{eq:fourier-noise} by the sample average over calibration bitstrings $\{\pmb{s}^{(k)}\}_{k=1}^{M_{\text{c}}}$,
\begin{equation}
    \hat{g}(\pmb{w}) = \frac{1}{M_c} \sum_{k=1}^{M_c}  (-1)^{\dprod{\pmb{w}}{\pmb{s}^{(k)}}}.
    \label{eq:gw-estimate}
\end{equation}
Note that in order to compute the sampled shades~\eqref{eq:mitigated-shade},  actually $1/\hat{g}(\pmb{w})$ is used. Using the Hoeffding inequality~\cite{Hoeffding1963a}, we show in Appendix~E that in order to guarantee
$\Pr \{ \abs{1/\hat{g}(\pmb{w}) - 1/g(\pmb{w})} \ge \epsilon \} \le \delta$, 
one requires
\begin{equation}
    M_{\text{c}} > -  32 \ln (\delta/ 2) /\epsilon^2 \times 1/g^{2} (\pmb{w}).
    \label{eq:calib}
\end{equation}

One sees that the number of samples requires to estimate $1/g (\pmb{w})$ scales with $1/g^{2} (\pmb{w})$. 
If one takes the simple model of independent-flips for the readout noise, it can be easily shown that that $g(\pmb{w})$ decays exponentially in the Hamming norm of the wavevector $\pmb{w}$; see Appendix E.  
Fourier components of higher norm are thus harder to estimate. 
This behaviour is also generally expected even in the present of crosstalk, as illustrated in our simulation below.

With the calibration data, the shadow tomography can be carried out as follow:

\smallskip
\begin{tabular}{l}
\textbf{$X$-twirled shadow tomography} \\
\hline
\hline
Repeat the following steps $M_{\text{st}}$ times: \\
1. Prepare the system in quantum state $\rho$ \\

2. Draw random directions $\pmb{\nu}=\{\vec{\nu}_i\}$ and apply the  \\  corresponding unitaries $G^{(\vec{\nu}_i)}$ on the qubits \\

3. Draw a random bitstring $\pmb{t}=\{t_i\}$ \\

4. Apply an $X$ gate at qubit $i$ if $t_i=1$ \\

5. Measure the system in the computational basis \\

6. Flip the obtained bit $i$ if $t_i=1$ \\

7. Record the final generalised outcome $(\pmb{\nu},\pmb{s})$ \\
\hline
\hline
\textbf{Output:} Generalised outcomes $\{(\pmb{\nu}^{(l)},\pmb{s}^{(l)})\}_{l=1}^{M_{\text{st}}}$
\end{tabular}
\smallskip

A correlation function $c^{\pmb{\mu}}_{\pmb{v}}= \mean{C^{\pmb{\mu}}_{\pmb{v}}}$ is then estimated by averaging~\eqref{eq:mitigated-shade} over the sampled generalised outcomes,
\begin{equation}
    \hat{c}^{\pmb{\mu}}_{\pmb{v}} = \frac{1}{M_{\text{st}}} \sum_{l=1}^{M_{\text{st}}}
    \frac{(-1)^{\dprod{\pmb{v}}{\pmb{s}^{(l)}}}}{\hat{g}(\pmb{v})} \prod_{i=1}^{n}  [\delta_{v_i,0}  +  \delta_{v_i,1}\frac{1}{2} \tr(\xi^{\vec{\nu}_i^{(l)}} \sigma^{\vec{\mu}_i})].
\end{equation}
Using the Hoeffding inequality, we show in Appendix E that  in order to guarantee 
$\operatorname{Pr}\{\abs{\hat{c}^{\pmb{\nu}}_{\pmb{s}} - c^{\pmb{\nu}}_{\pmb{s}}} \ge \epsilon \} \le \delta$,
it is required that
\begin{equation}
    M_{\text{st}} > -2 \ln (\delta/2)/\epsilon^2 \times   \kappa^{2\abs{v}}/g^{2}(\pmb{w}),
    \label{eq:st}
\end{equation}
where $\kappa = 1/2 \max \{\tr (\xi^{\vec{\nu}_i} \sigma^{\vec{\mu}_i}) \}$.
One also observes that the factor $\kappa^{2\abs{v}}$ contributes an exponential scaling of the required  number of samples for shadow tomography. 
This is known in shadow tomography~\cite{huang_predicting_2020}.
As regarding the effect of the readout noise, the required number of samples scales as $1/g^{2}(\pmb{w})$, similarly to the complexity of the calibration process.

\textit{Simulation using random circuits---} To illustrate how the protocol works in practice, we 
carry out a simulation of a quantum computer backend of $n=8$ qubits. 
The readout noise is included for each qubit with flipping rate of $7\%$ from $1$ to $0$, and $5\%$ from $0$ to $1$. The crosstalk is then included using the model suggested by Ref.~\cite{Li2023a}.

\begin{figure}[t]
    \centering
    \includegraphics[width=0.48\textwidth]{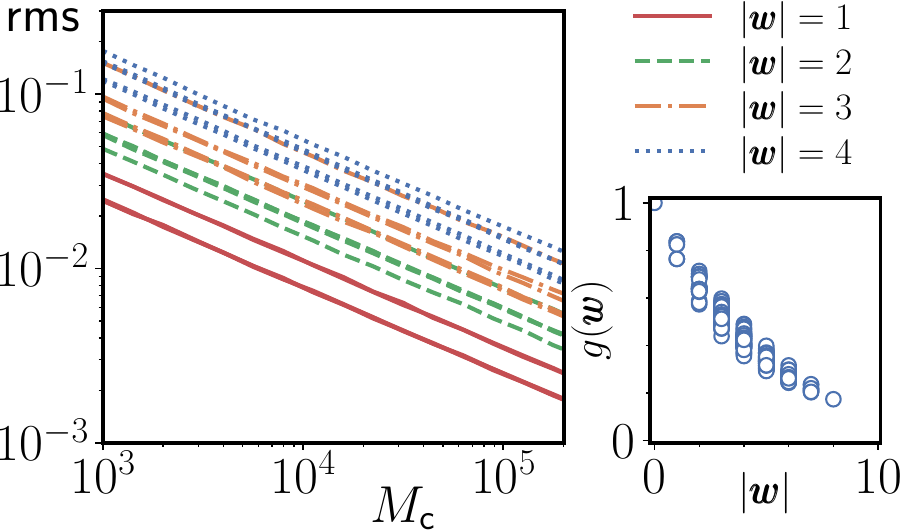}
    \caption{(left) The root mean squared errors (rms) of $1/\hat{g} (\pmb{w})$ decreases as $M_{\text{c}}^{-1/2}$. Estimation of $1/g (\pmb{w})$ is harder as the wavenumber $\abs{\pmb{w}}$ increases. (right) The Fourier component $g (\pmb{w})$ decreases exponentially as the wavenumber $\abs{\pmb{w}}$ increases. A basis dataset of $10^7$ calibration data points was collected, and bootstrap with replacement was subsequently used to simulate datasets at different sampling sizes.}
    \label{fig:calibration}
\end{figure}

Figure~\ref{fig:calibration} (left) illustrates the convergence of the inverse of the estimated Fourier component $1/g(\pmb{w})$ with increasing calibration data $M_{\text{c}}$ collected. 
One observes the decreasing of the root mean squared errors ($\text{rms}$) as $M_{\text{c}}^{-1/2}$ as expected from equation~\eqref{eq:calibration-complexity}. 
The lower right panel of Figure~\ref{fig:calibration} also illustrates that the Fourier component $g(\pmb{w})$ decays approximately exponentially with respect to the wavenumber $\abs{\pmb{w}}$. 
This implies directly by equation~\eqref{eq:calibration-complexity} that the inverse of the Fourier components of higher wavevectors are harder to estimate, as also seen in the left panel. 

\begin{figure}[!t]
    \centering
    \includegraphics[width=0.48\textwidth]{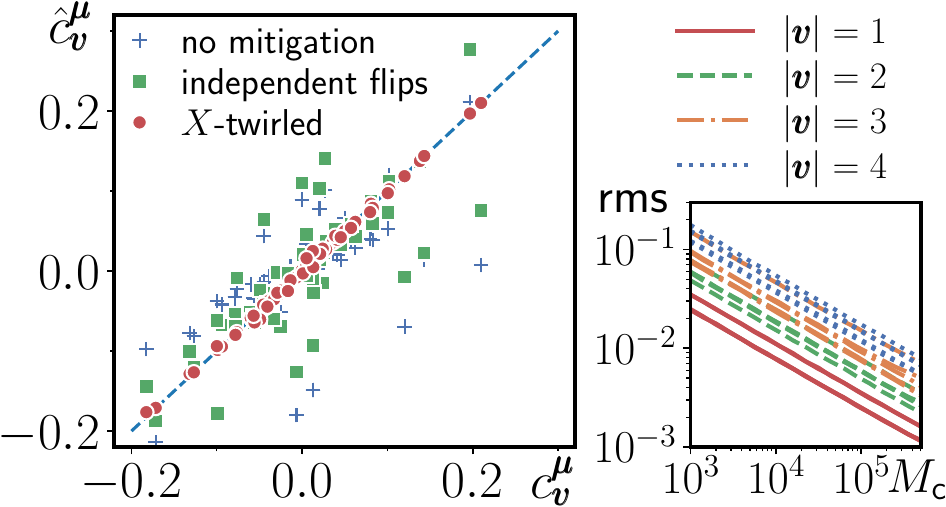}
    \caption{(left) Reconstructed correlation functions from shadow tomography with noisy readout using no mitigation, mitigation with independent-flip model and the method of $X$-twirling.  (right) Root mean square errors (rms) of the estimator for the correlation function $\hat{c}^{\pmb{\mu}}_{\pmb{v}}$ by $X$-twirling decreases as $M_{\text{st}}^{-1/2}$. Estimation of the correlation function $c^{\pmb{\mu}}_{\pmb{v}}$ is harder as $\abs{\pmb{v}}$ increases. A basis dataset of $10^7$ shadow tomography data points was collected, and bootstrap with replacement was used to simulate datasets at different sampling sizes.}
    \label{fig:shadow-tomography}
\end{figure}

To investigate the performance of shadow tomography, we simulate a random circuit of depth $20$ for state preparation. 
The obtained state is then subject to shadow tomography with the random directions for measurements in $x$, $y$ and $z$. 
We then randomly select correlators of degree $|\pmb{v}|=1,2,3,4$ also with random Pauli observables and compute the corresponding correlation functions. 
For comparison, we also carry out shadow tomography without mitigation of readout noise~\cite{huang_predicting_2020}, and with mitigation of readout noise under the assumption that outcome bits are independently flipped using the method of generalised measurements~\cite{Nguyen2022a}. 
Figure~\ref{fig:shadow-tomography} (left) shows that without correction or with the independent-flips model, the reconstructed correlation functions by shadow tomography can contain significant bias from the true values. 
In contrast, $X$-twirling allows for accurate estimation of the correlation without systematic bias. 
Figure~\ref{fig:shadow-tomography} (right) demonstrates the convergence of the estimated correlation functions as the collected shadow tomography data increases. One again observes that the root mean squared errors of the estimators of the correlation functions decrease as $M_{\text{st}}^{-1/2}$. That correlators of higher degree  $|\pmb{v}|$ are harder to estimate is again clearly illustrated.

\textit{Conclusion---} We demonstrate that shadow tomography can be carried out reliably on devices with complicated readout noise using the simple method of $X$-twirling. 
This was a surprising consequence of not only of the $X$-twirled noise, but also the specific definition of noisy classical shadows and their Fourier presentation. 
It would be interesting to clarify whether $X$-twirling is also sufficient for other shadow tomography schemes, such as the mutual unbiased shadow tomography~\cite{Wang2023a}, or fermionic shadow tomography~\cite{Zhao2021a}.
As for more sophisticated shadow tomography schemes, even when $X$-twirling might turn out be not sufficient, further twirling with qubit swapping is very promising. 
Another important direction to be explored is how nonlinear properties such as the entropy can be estimated from the noisy shadow tomography.
In this case, the integration of methods of readout error mitigation that can avoid negative probabilities such as the newly rediscovered iterative bayesian unfolding~\cite{nachman_unfolding_2020,srinivasan_scalable_2022,Nguyen2023a} is also important.

\begin{acknowledgments}
The author would like to thank 
Otfried G\"uhne,
Matthias Kleinmann,
Yi Li, 
and
Jonathan Steinberg, 
for helpful discussions.
Comments from Tung-Lam Nguyen and Thi-Trang Le
were encouraging. 
The University of Siegen is kindly acknowledged for enabling our computation through the \texttt{OMNI} cluster. 
This work was supported by 
the Deutsche Forschungsgemeinschaft (DFG, German Research Foundation, project numbers 447948357 and 440958198), 
the Sino-German Center for Research 
Promotion (Project M-0294), 
the ERC (Consolidator Grant 683107/TempoQ), 
and the German Ministry of Education 
and Research (Project QuKuK, BMBF 
Grant No. 16KIS1618K).
\end{acknowledgments}

\appendix

\section{Noiseless shadow tomography}
\label{sec:noiseless}
As mentioned in the main text, we concentrate on shadow tomography with the separable qubit measurement scheme as illustrated in Fig.~\ref{fig:noisy-st}a, which is arguably most practical for the available quantum simulators due to its simplicity. 
In this scheme,  each of the $n$ qubits of the quantum simulator prepared in the desired state is measured in a random direction given by the unit vectors $\pmb{\nu}=\{\vec{\nu}_i\}_{i=1}^{n}$ choosing among a set $\mathcal{S}$ of predefined directions. 
As a typical example, the measurements can be chosen to be Pauli measurements in directions $\mathcal{S}=\{x,y,z\}$; our analysis is however not limited to this particular case. 
In practice, instead of rotating the measurement devices, one applies an appropriate local unitary gate $G^{(\vec{\nu}_i)}$ to qubit $i$ such that the measurement in the computational basis corresponds to the measurement in the desired direction $\vec{\nu}_i$ with projections $P_{s_i}^{\vec{\nu}_i} = \ketbra{\vec{\nu}_i,s_i}{\vec{\nu}_i,s_i}$, that is,
\begin{equation}
    P_{s_i}^{\vec{\nu}_i} = [G^{(\vec{\nu}_i)}]^{\dagger} P_{s_i}^{z} G^{(\vec{\nu}_i)},
    \label{eq:def-U}
\end{equation}
where $P_{s_i}^{z}$ with $s_i=0$ and $s_i=1$ are the two projections onto the computational basis of qubit $i$.
Each of such measurement yields a bitstring $\pmb{s}=\{s_i\}_{i=1}^{n}$ as the outcome.
The pair $(\pmb{\nu},\pmb{s})$ for each run is then referred to as a \emph{generalised outcome}.

In fact, it is most convenient to view this whole randomisation procedure as a single generalised measurement making on the qubits~\cite{Nguyen2022a}. Without noise, this generalised measurement is given by  effects $P^{\pmb{\nu}}_{\pmb{s}}$ characterising every generalised outcome $(\pmb{\nu},\pmb{s})$ of all $\Omega=\abs{\mathcal{S}}^n \times 2^n $ possibilities. Explicitly, the effects are given by
\begin{equation}
    P^{\pmb{\nu}}_{\pmb{s}} = \frac{1}{\abs{\mathcal{S}}^n} \otimes_{i=1}^{n} P^{\vec{\nu}_i}_{s_i},
\end{equation}
where $P^{\vec{\nu}_i}_{s_i}$ is the projection of the qubit onto state $s_i$ in direction $\vec{\nu}_i$ as in equation~\eqref{eq:def-U}.
Such a generalised measurement can be considered as a hermitian map $\Phi_0: \CC^{2^n \times 2^n} \to \RR^{\Omega}$, which maps a density operator $\rho$ to a classical distribution over generalised outcomes $(\pmb{\nu},\pmb{s})$, $p= \Phi_0 [\rho]  \in \RR^{\Omega}$, explicitly given by
\begin{equation}
    p({\pmb{\nu}},{\pmb{s}}) = \tr(\rho P^{\pmb{\nu}}_{\pmb{s}}).
    \label{eq:noiseless-measurement-channel}
\end{equation}
The generalised outcome $(\pmb{\nu},\pmb{s})$ can thus be understood as a random variable distributed according to the distribution $p({\pmb{\nu}},{\pmb{s}})$.
Each of the run of the protocol gives then one sample of this random variable. 
After $L$ runs, one obtains a dataset $\{(\pmb{\nu}^{(l)},\pmb{s}^{(l)})\}_{l=1}^{L}$ of $L$ such samples, which is the basic data that capture the behaviour of the (unknown) quantum state. 
Notice that for systems of few dozens of qubits, the dataset $\{(\pmb{\nu}^{(l)},\pmb{s}^{(l)})\}_{l=1}^{L}$ is typically rather dilute, $L \ll \Omega=\abs{\mathcal{S}}^n \times 2^n $. 

From this sampling data, we would like to have an estimate for the original density operator $\rho$. An unbiased linear estimator for the density operator $\rho$ is a map $\chi_0: \RR^{\Omega} \to \CC^{2^n \times 2^n}$ such that $\Phi_0 \circ \chi_0$ acts as the identity in $\RR^{\Omega}$.  
The canonical choice for the estimator would be the least-square estimator~\cite{guta_fast_2020,Nguyen2022a}, explicitly given by
\begin{equation}
\chi_{0} = (\Phi_0^{\dagger} \Phi_0)^{-1} \Phi_0^{\dagger}.
\end{equation}
This leads to the estimated state as
\begin{equation}
    \hat{\rho} = \chi_0 (\hat{p}),
    \label{eq:formal-estimator}
\end{equation}
where $\hat{p}$ is the sampled distribution constructed from the dataset $\{(\pmb{\nu}^{(l)},\pmb{s}^{(l)})\}_{l=1}^{L}$,
\begin{equation}
    \hat{p}(\pmb{\nu},\pmb{s}) = \frac{1}{L}\sum_{l=1}^{L}  \delta_{\pmb{\nu},\pmb{\nu}^{(l)}} \delta_{\pmb{s},\pmb{s}^{(l)}}.  
\end{equation}
In scalable quantum information processing, where the system size $n$ is large, the density operator $\rho$ is exponentially large and its explicit computation by means of~\eqref{eq:formal-estimator} is infeasible.
Shadow tomography starts with the idea that even when the estimated density operators $\hat{\rho}$ cannot be written down explicitly, most often the mean values of an observable can still be computed efficiently~\cite{aaronson_journal_2020,huang_predicting_2020}.

One first observes that by the linearity of $\chi_0$, the estimate for the density operator $\hat{\rho}$ can be represented as a sum over sampled \emph{classical shadows},
\begin{equation}
    \hat{\rho} = \frac{1}{L}\sum_{l=1}^{L} \rho^{\pmb{\nu}^{(l)}}_{\pmb{s}^{(l)}} ,
\end{equation}
where the classical shadows are defined as the estimated density operator for a delta distribution defined by a single outcome, $\rho^{\pmb{\nu}^{(l)}}_{\pmb{s}^{(l)}} = \chi_0 [\{\delta_{\pmb{\nu},\pmb{\nu}^{(l)}} \delta_{\pmb{s},\pmb{s}^{(l)}}\}]$.

In other words, every single generalised outcome $(\pmb{\nu},\pmb{s})$ is associated with a delta distribution in $\RR^{\Omega}$, which in turn can be turned into a classical shadow $\rho^{\pmb{\nu}}_{\pmb{s}}$ by the estimator $\chi_0$. 
Accordingly, one transforms from the dataset of (dilute) sampled generalised outcomes $\{(\pmb{\nu}^{(l)},\pmb{s}^{(l)} )\}_{l=1}^{L}$ to a dataset of (dilute) sampled classical shadows $\{\rho^{\pmb{\nu}^{(l)}}_{\pmb{s}^{(l)}}\}_{l=1}^{L}$, the average of which corresponds to the estimate of $\rho$. 
In particular, in the infinite sampling limit, 
\begin{equation}
   \lim_{L \to \infty} \frac{1}{L} \sum_{l=1}^{n} \rho^{\pmb{\nu}^{(l)}}_{\pmb{s}^{(l)}} = \sum_{{\pmb{\nu}},{\pmb{s}}} \rho^{\pmb{\nu}}_{\pmb{s}} \tr (\rho P^{\pmb{\nu}}_{\pmb{\sigma}})
   \label{eq:naive-limit}
\end{equation}
converges to $\rho$ due to the choice of the unbiased estimator in defining the classical shadows.

Shadow tomography hings on the fact that by certain design of the measurements, the classical shadows can be efficiently presented~\cite{aaronson_journal_2020,huang_predicting_2020}. 
Further, they can then be efficiently transformed to a the dataset of sampled mean values for a given observable.
In particular, for measurements making on single qubits, the measurement channel $\Phi_0$ factorises over the qubits. 
As a result, the estimator $\chi_0$ also factorises over the qubits. 
Eventually, each of the classical shadows $\rho_{\pmb{\nu}^{(l)}, \pmb{s}^{(l)}}$ can be \emph{efficiently} presented by a \emph{tensor product} over the qubits~\cite{huang_predicting_2020}, 
\begin{equation}
    \rho^{\pmb{\nu}}_{\pmb{s}} = \otimes_{i=1}^{n} \rho_{s_i}^{\vec{\nu}_i}.
    \label{eq:noiseless-shadow-real}
\end{equation}
The single qubit classical shadows $\rho_{s_i}^{\vec{\nu}_i}$ can be generically written as 
\begin{equation}
    \rho_{s_i}^{\vec{\nu}_i} = \frac{1}{2}[\openone + (-1)^{s_i} \xi^{\vec{\nu}_i}]
    \label{eq:noiseless-shadow-real-1}
\end{equation}
for certain traceless operators $\xi^{\vec{\nu}_i}$. Observe that the classical shadow $\rho_{s_i}^{\vec{\nu}_i}$ depends on $s_i$ in a very particular way, indicating the inversion symmetry inherited from the symmetry between two projections for each measurement in direction $\vec{\nu}_i$. This  turns out to be crucial to the ability to mitigate readout errors for classical shadows. 

To see this symmetry formally, consider the generalised measurement restricted to a single qubit, described by effects $\{1/\abs{\mathcal{S}} \ketbra{\vec{\nu}_i,s_i}{\vec{\nu}_i,s_i} \}_{\vec{\nu} \in \mathcal{S}}$. This generalised measurement has an inherent symmetry in the sense that is generally defined in~\cite{nguyen_symmetries_2020} given by the conjugate action of the anti-unitary operator $\sigma_y T$, $\sigma_y T \ketbra{\vec{\nu}_i,s_i}{\vec{\nu}_i,s_i} T \sigma_y = \ketbra{\vec{\nu}_i,s_i}{\vec{\nu}_i,s_i}$, where $T$ is the complex conjugate operator. Geometrically, this simply reflects the inversion symmetry of the Bloch sphere. It has been already remarked that the symmetry of the measurement immediately implies the symmetry of the classical shadows constructed by means of least-square estimator~\cite{Nguyen2022a}. Therefore the classical shadows also distribute centrally symmetric in the state space.

It should be noted that the single qubit classical shadows $\rho_{s_i}^{\vec{\nu}_i}$ and thus also $\rho^{\pmb{\nu}}_{\pmb{s}}$ are not necessarily positive.
As an example, for shadow tomography with ideal Pauli measurements, $\xi^{\alpha} = 3 \sigma^{\alpha}$ with  $\sigma^{\alpha}$ being the Pauli observables for $\alpha=x,y,z$~\cite{huang_predicting_2020}.

Most often, one is interested in estimating correlation functions from the simulated quantum  state. 
Such a correlation function is the expectation value of an observable acting over a group of few qubits, hereafter called correlation operator, or correlator as in the main text. 
To indicate the group of interested qubits for a correlator, one can use a bitstring $\pmb{v}= \{v_i\}$ where $v_i=1$ indicates its non-trivial action on qubit $i$. 
The vector $\pmb{v}$ is referred to as the pattern of the correlator.
Notice that the Hamming norm of $\pmb{v}$, denoted by $\abs{\pmb{v}}$, indicates the number of qubits where the correlator acts non-trivially. In the literature, $\abs{\pmb{v}}$ is called the locality of the observable; it is perhaps better called the degree of the correlator. 
Then a correlator is further fully specified by a set of directions $\pmb{\mu}=\{\vec{\mu}_i\}$ indicating the single qubit observables.  
Explicitly, a correlator
$C^{\pmb{\mu}}_{\pmb{v}}$ can be written as
\begin{equation}
    C^{\pmb{\mu}}_{\pmb{v}} = \otimes_{i=1}^{n} [\delta_{v_i,0} \openone + \delta_{v_i,1} \sigma^{\vec{\mu}_i}],
\end{equation}
with $\sigma^{\vec{\mu}_i}= \vec{\mu}_i \cdot \vec{\sigma}$ being the qubit observable defined by direction $\vec{\mu}_i$, where $\vec{\sigma} = (\sigma^x,\sigma^y,\sigma^z)$ . 

The corresponding correlation function, $c^{\pmb{\mu}}_{\pmb{v}} = \tr [\rho C^{\pmb{\mu}}_{\pmb{v}}]$, is then estimated by $\hat{c}^{\pmb{\mu}}_{\pmb{v}} = \tr [\hat{\rho} C^{\pmb{\mu}}_{\pmb{v}}]$. Using the classical shadow presentation, one can present the estimator $\hat{c}^{\pmb{\mu}}_{\pmb{v}}$ of the expectation value  as the (sample) average over a random variable
\begin{equation}
    \hat{c}^{\pmb{\mu}}_{\pmb{v}} = \frac{1}{L}\sum_{l=1}^{L} c^{\pmb{\mu},\pmb{\nu}^{(l)}}_{\pmb{v},\pmb{s}^{(l)}}
\end{equation}
where 
\begin{equation}
    c^{\pmb{\mu},\pmb{\nu}}_{\pmb{v},\pmb{s}} = \tr [\rho^{\pmb{\nu}}_{\pmb{s}} C^{\pmb{\mu}}_{\pmb{v}}], 
\end{equation}
which is called the \emph{shade} for correlator $C^{\pmb{\mu}}_{\pmb{v}}$ formed by the classical shadow $\rho^{\pmb{\nu}}_{\pmb{s}}$.

Crucially, the shade $c^{\pmb{\mu},\pmb{\nu}}_{\pmb{v},\pmb{s}}$ for a correlator can be efficiently computed as long as the classical shadows factorise, 
\begin{equation}
    c^{\pmb{\mu},\pmb{\nu}}_{\pmb{v},\pmb{s}} = \frac{1}{L} \sum_{l=1}^{L} \prod_{i=1}^{n}[\delta_{v_i,0} + (-1)^{s_i^{(l)}} \delta_{v_i,1} \frac{1}{2} \tr (\xi^{\vec{\nu}_i^{(l)}} \sigma^{\vec{\mu}_i}) ].
    \label{eq:snapshot-mean}
\end{equation}
It is again to be emphasised that the right hand side of~\eqref{eq:snapshot-mean} can be efficiently evaluated without computing the estimator for the large density operator $\hat{\rho}$ explicitly. 
This is carried out by first transforming the dataset of sampling the random generalised outcome $(\pmb{\nu},\pmb{s})$ to a dataset of sampling classical shadows $\rho^{\pmb{\nu}}_{\pmb{s}}$, which is then transformed directly into sampling shades $c^{\pmb{\mu},\pmb{\nu}}_{\pmb{v},\pmb{s}}$ for the correlation function of interest. 

It is then clear that starting from the sampling data  $\{(\pmb{\nu}^{(l)},\pmb{s}^{(l)})\}_{l=1}^{L}$, any correlation function can be estimated in a computationally efficient way. 
In principle, several observables can also be \emph{simultaneously} estimated in this way.
However,  constraining the simultaneous accuracy of many expectation values by the sample mean estimator can require higher sample complexity due to the simultaneous fluctuation of the estimators. 
The problem can be resolved by using the median-of-means estimator~\cite{huang_predicting_2020}.
In this way, simultaneous fluctuation of the estimators for the means is suppressed, allowing for high accuracy estimation of them jointly.
In this sense, shadow tomography can be considered as a rather faithful description of the quantum state to be characterised.
For the details of the method of median of means applied to shadow tomography, we refer the readers to Ref.~\cite{huang_predicting_2020}.

\section{Shadow tomography with noisy readout}
\label{sec:noisy}

So far we have assumed that the quantum simulator works perfectly without any noise.
Unfortunately, for the available and near-future devices, this is too strong an assumption. 
Indeed, for the available devices, not only the gates cannot be perfectly implemented, but also the readout of the measurements is far from perfect. 

As we mentioned in the main text that gate imperfection would affect not only the realisation of the random directions for the measurements in the shadow tomography protocol, but also the quality of the state prepared by the simulator. 
Mitigation of gate errors for state preparation can be then naturally extended to include random gate required by shadow tomography. 

On other hand, readout errors directly influence the data underlying shadow tomography.
In that sense, it is desirable that mitigation of readout errors is addressed within the shadow tomography protocol.

As we mentioned in the main text, readout errors are typically modelled by a two-step process.
In the actual measurement, it is supposed that the ideal, exact outcome $\pmb{s}'$ of the measurement are generated, but cannot be observed.
The actual observed outcome that is readout $\pmb{s}$ is obtained by altering $\pmb{s}'$ with certain transition probability $R(\pmb{s} \vert \pmb{s}')$, which depends on the details of the setup of the device.
While this model may not capture all the physics of readout errors, it proves to be useful and has become the arguably most widely used model for readout errors for superconducting quantum computers~\cite{Chen2019}.

Suppose for a given measurement setting $\pmb{\nu}$, the ideal, unobservable outcome $\pmb{s}'$ has a distribution $p(\pmb{\nu},\pmb{s}')$, the actual registered distribution $q(\pmb{\nu},\pmb{s}')$ is then
\begin{equation}
    q(\pmb{\nu},\pmb{s}) = \sum_{\pmb{s}'} R (\pmb{s} | \pmb{s}') p(\pmb{\nu},\pmb{s}').
    \label{eq:transition-equation}
\end{equation}
Effectively, this leads to the replacement of the noiseless generalised measurement effects $P^{\pmb{\nu}}_{\pmb{s}}$ by the noisy ones $E^{\pmb{\nu}}_{\pmb{s}}$ given by
\begin{equation}
    E_{\pmb{s}}^{\pmb{\nu}} = \sum_{\pmb{s}'} R (\pmb{s} | \pmb{s}') P_{\pmb{s}'}^{\pmb{\nu}}.
    \label{eq:noisy-effect}
\end{equation}
The ideal measurement channel $\Phi_0$ in equation~\eqref{eq:noiseless-measurement-channel} is also replaced by the noisy measurement channel, $\Phi: \CC^{2^n \times 2^n} \to \RR^{\Omega}$,
\begin{equation}
    \Phi [\rho] (\pmb{\nu},{\pmb{s}}) = \tr(\rho E^{\pmb{\nu}}_{\pmb{s}}).
\end{equation}
As before, an unbiased linear estimator is a map $\chi: \RR^{\Omega} \to \CC^{2^n \times 2^n}$ so that $\Phi \circ \chi$ acts as the identity. 
One may attempt to use the least square construction for the estimator. 
This choice together with the assumption that the outcome bits are flipped independently at readout, that is 
\begin{equation}
    R (\pmb{s} | \pmb{s}') = \prod_{i=1}^{n} R_i(s_i | s_i'),
    \label{eq:factorised-noise}
\end{equation}
for single qubit transition rates $R_i(s_i | s_i')$,
indeed leads to a simple error mitigation of readout noise for shadow tomography~\cite{Nguyen2022a}.

Unfortunately, more and more recent investigations indicated that the independent-flip model~\eqref{eq:factorised-noise} is too simplistic.
It is suggested that crosstalk in readout errors can be significant in certain devices~\cite{Bravyi2021a,Berg2022a}.  
This demands for scalable mitigation technique for classical shadows with consideration of readout crosstalk.
As one might already expect, once the crosstalk in the readout errors is significant, the resulted classical shadows from the least-square estimator generally do not admit an efficient presentation in the form of tensor product over many qubits such as~\eqref{eq:noiseless-shadow-real}. The classical shadow information processing pipeline thus breaks down for large systems.

To resolve this problem, one has to abandon canonical choices for classical shadows. 
We suggest in the main text the following classical shadows for the data with readout noise
\begin{equation}
    \tilde{\rho}^{\pmb{\nu}}_{\pmb{s}} =  \sum_{\pmb{s}'} \rho^{\pmb{\nu}}_{\pmb{s}'} R^{-1} (\pmb{s}' |\pmb{s}),
    \label{eq:noisy-shadow-real}
\end{equation}
where $\rho^{\pmb{\nu}}_{\pmb{s}'}$ are the noiseless classical shadows in equation~\eqref{eq:noiseless-shadow-real} and $R^{-1}$ denotes the inverse of the transition probabilities as a matrix. 
Notice the order of the arguments of $R^{-1}$.
That this gives rise to an unbiased estimator can be directly proven by
\begin{align}
    \sum_{\pmb{\nu},\pmb{s}} \tilde{\rho}^{\pmb{\nu}}_{\pmb{s}} q (\pmb{\nu},\pmb{s}) & = \sum_{\pmb{\nu},\pmb{s}} \sum_{\pmb{s'}} \rho^{\pmb{\nu}}_{\pmb{s}'} R^{-1} (\pmb{s}' |\pmb{s})  \sum_{\pmb{s}''} R (\pmb{s} | \pmb{s}'') p(\pmb{\nu},\pmb{s}'') \nonumber \\
    & = \sum_{\pmb{\nu},\pmb{s}'}  \rho^{\pmb{\nu}}_{\pmb{s}'}  p(\pmb{\nu},\pmb{s}'),
\end{align}
where the last expression converges to the state $\rho$ as the fundamental property of the noiseless classical shadows~\eqref{eq:naive-limit}. Convergence of mean values of observables thus follows.

The definition of the classical shadows~\eqref{eq:noisy-shadow-real} is however impractical. Indeed, it involves the inversion of the exponentially large transition matrix $R(\pmb{s} | \pmb{s}')$. In fact, the transition matrix $R(\pmb{s} | \pmb{s}')$ itself cannot be estimated for large systems because of its exponential size.
Fortunately, by introducing $X$-twirling, i.e., random flips before and after the measurement in the computational basis, $R(\pmb{s} | \pmb{s}')$ is translationally symmetrised. In this case, the classical shadows~\eqref{eq:noisy-shadow-real} accept an efficient presentation in \emph{Fourier space}, which again facilitates efficient computation of correlation functions in the same way as ideal classical shadows do.

\section{Twirling and translationally symmetric noise}

As we mentioned in the main text, the idea of $X$-twirling for readout noise~\cite{Berg2022a} is to pick a random bitstring $\pmb{t}$ in $\{0,1\}^n$ and apply an $X$-gate to qubit $i$ if $t_i=1$ prior to the measurement. The obtained outcome bitstring is then (classically) flipped back at the same position $i$.
Effectively, this allows one to replace the original noise with the twirled one,
\begin{equation}
    \bar{R}(\pmb{s} | \pmb{s}') = \frac{1}{2^n} \sum_{\pmb{t}} R(\pmb{s} \oplus \pmb{t} | \pmb{s}' \oplus \pmb{t}),
    \label{eq:twirled-1}
\end{equation}
where $\oplus$ denotes the summation of bits modulo $2$.

To see this, observe that the probability of observing an outcome $\pmb{s}$  with a given selection of the random flips $\pmb{t}$ is 
\begin{equation}
    q(\pmb{\nu},\pmb{s} | \pmb{t}) = \sum_{\pmb{s}'} R(\pmb{s} \oplus \pmb{t} | \pmb{s'} ) \tr [X^{\pmb{t}} G^{(\pmb{\nu})} \rho (G^{(\pmb{\nu})})^{\dagger} X^{\pmb{t}}  P_{\pmb{s}'}^{\pmb{z}}].
\end{equation}
Here we use $X^{\pmb{t}} = \otimes_{i=1}^{n} X^{t_i}$, $G^{(\pmb{\nu})} = \otimes_{i=1}^{n} G^{(\vec{\nu}_i)}$ with the gates $G^{(\vec{\nu}_i)}$ defined in equation~\eqref{eq:def-U}.

Notice then that $\tr [X^{\pmb{t}} G^{(\pmb{\nu})} \rho (G^{(\pmb{\nu})})^{\dagger} X^{\pmb{t}}  P^{\pmb{\nu}}_{\pmb{s}'}] = \tr [\rho (G^{(\pmb{\nu})})^{\dagger} X^{\pmb{t}}  P^{\pmb{z}}_{\pmb{s}'} X^{\pmb{t}} G^{(\pmb{\nu})}]$, and $(G^{(\pmb{\nu})})^{\dagger} X^{\pmb{t}}  P^{\pmb{z}}_{\pmb{s}'} X^{\pmb{t}} G^{(\pmb{\nu})} = P^{\pmb{\nu}}_{\pmb{s}' \oplus \pmb{t}} $, we arrive at 
\begin{equation}
    q(\pmb{\nu},\pmb{s} | \pmb{t}) = \sum_{\pmb{s}'} R(\pmb{s} \oplus \pmb{t} | \pmb{s'} \oplus \pmb{t}  ) \tr [\rho   P^{\pmb{\nu}}_{\pmb{s}'}  ].
\end{equation} 
Assuming the bitstring $\pmb{t}$ is chosen at random, by marginalising over $\pmb{t}$, one obtains,
\begin{equation}
    q(\pmb{\nu},\pmb{s}) = \sum_{\pmb{s}'} \frac{1}{2^n} \sum_{\pmb{t}} R(\pmb{s} \oplus \pmb{t} | \pmb{s'} \oplus \pmb{t}  ) \tr [\rho   P^{\pmb{\nu}}_{\pmb{s}'}  ].
    \label{eq:twirled-2}
\end{equation}
One sees then that equation~\eqref{eq:twirled-2} assumes the same form as equation~\eqref{eq:transition-equation} with the twirled noise given by~\eqref{eq:twirled-1}. 

The twirled transition matrix $\bar{R}(\pmb{s} | \pmb{s}')$ has an important property that it is translationally invariant, 
\begin{equation}
    \bar{R}(\pmb{s} | \pmb{s}') = \bar{R}(\pmb{s} \oplus \pmb{t} | \pmb{s}' \oplus \pmb{t}),
\end{equation}
which will be crucial for our mitigation protocol.
In particular, we have
\begin{equation}
    \bar{R}(\pmb{s} | \pmb{s}') = \bar{R}(\pmb{s} \oplus \pmb{s}' | \pmb{0}).
\end{equation}
One observes that in order to estimate the original transition matrix $R (\pmb{s} | \pmb{s}')$, one has to prepare $2^n$ states $\ket{\pmb{s}'}$ by constructing the corresponding circuits, which are subjected to repeated measurements to estimate the frequencies of the observed outcomes $\pmb{s}$. In contrast, for a symmetric transition matrix $\bar{R} (\pmb{s} \oplus \pmb{s}' | \pmb{0})$, it is sufficient to prepare a single state $\ket{\pmb{0}}$ and repeatedly measure in the computational basis. In principle, there are still $2^n$ frequencies to be estimated. However, the frequencies for frequent outcomes can be reasonable estimated while rare outcomes can be ignored without significantly affecting the accuracy in downstream information processing.

\section{Mitigation of classical shadows under translationally symmetric noise}

Formally, the space of bitstrings $\{0,1\}^n$ can be considered as a linear vector space of $n$-dimension over the field $\{0,1\}$. In particular, $\{0,1\}^n$ acts on itself as a linear translational group, $T_{\pmb{t}}: \Omega \to \Omega$, $T_{\pmb{t}} (\pmb{s}) = \pmb{s}  \oplus \pmb{t}$ for $\pmb{t} \in \{0,1\}^n$. 

Importantly, in the space $\{0,1\}^n$ one also has a Fourier kernel given by $(-1)^{\dprod{\pmb{w}}{\pmb{s}}}$ with the identity $\sum_{\pmb{s}} (-1)^{\dprod{\pmb{w}}{\pmb{s}}} = 2^n \delta_{\pmb{w},\pmb{0}}$. 
While this Fourier kernel has somewhat peculiar form, it mimics the familiar Fourier transform  in physics in many aspects.
In particular, as the transition matrix $\bar{R} (\pmb{s}|\pmb{s}')$ is translationally invariant, it is \emph{diagonal} in the Fourier space. 
Indeed, one can write
\begin{equation}
    \bar{R} (\pmb{s}|\pmb{s}') = \frac{1}{2^n}\sum_{\pmb{w}} (-1)^{\dprod{\pmb{w}}{\pmb{s} \oplus \pmb{s}'}} g(\pmb{w})
\end{equation}
where 
\begin{equation}
    g(\pmb{w}) = \sum_{\pmb{s}} (-1)^{\dprod{\pmb{w}}{\pmb{s}}} \bar{R} (\pmb{s} | \pmb{0}).
    \label{eq:def-gw}
\end{equation}
As such, the transition matrix is completely characterised by its Fourier components $g(\pmb{w})$. Notice that the definition~\eqref{eq:def-gw} is  in principle a sum over an exponentially large number of $\pmb{s}$. However, only few of the frequencies $\bar{R} (\pmb{s} | \pmb{0})$ are expected  to be significant, while most of them are practically zero. Therefore $g(\pmb{w})$ can be estimated by retaining only non-zero $\bar{R} (\pmb{s} | \pmb{0})$ observed in the calibration data. 

That $\bar{R} (\pmb{s}|\pmb{s}')$ has a simple Fourier presentation suggests that one can work more efficiently in the Fourier space. To this end, we define the Fourier transform of the noisy classical shadows 
\begin{equation}
    \tilde{\tau}^{\pmb{\nu}}_{\pmb{w}}  = \sum_{s} (-1)^{\dprod{\pmb{w}}{\pmb{s}}} \tilde{\rho}^{\pmb{\nu}}_{\pmb{s}},
    \label{eq:noisy-shadow-fourier}
\end{equation}
and noiseless classical shadows,
\begin{equation}
    {\tau}^{\pmb{\nu}}_{\pmb{w}}  = \sum_{s} (-1)^{\dprod{\pmb{w}}{\pmb{s}}} {\rho}^{\pmb{\nu}}_{\pmb{s}}.
\end{equation}
The relation~\eqref{eq:noisy-shadow-real} indeed becomes simple in the Fourier space,
\begin{equation}
    \tilde{\tau}^{\pmb{\nu}}_{\pmb{w}} = \frac{1}{g(\pmb{w})} {\tau}^{\pmb{\nu}}_{\pmb{w}}.
    \label{eq:noisy-noiseless-fourier}
\end{equation}

Using the unmitigated classical shadow $\rho^{\pmb{\nu}}_{\pmb{s}}$ in equation~\eqref{eq:noiseless-shadow-real}, one can easily compute its Fourier transform $\tau^{\pmb{\nu}}_{\pmb{w}}$ explicitly,
\begin{equation}
\tau_{\pmb{w}}^{\pmb{\nu}} = \otimes_{i=1}^{n} [\delta_{w_i,0} \openone + \delta_{w_i,1} \xi^{\vec{\nu}_i}].
\label{eq:noiseless-shadow-fourier-explicit}
\end{equation}
With~\eqref{eq:noiseless-shadow-fourier-explicit}, one can compute the mitigated classical shadows in the Fourier space $\tilde{\tau}^{\pmb{\nu}}_{\pmb{w}}$ using~\eqref{eq:noisy-noiseless-fourier}.
One then can compute the noisy classical shadow $\tilde{\rho}^{\pmb{\nu}}_{\pmb{s}}$ in real space by reverting its Fourier transform~\eqref{eq:noisy-shadow-fourier},
\begin{equation}
   \tilde{\rho}^{\pmb{\nu}}_{\pmb{s}} = \frac{1}{2^n} \sum_{\pmb{w}} (-1)^{\dprod{\pmb{w}}{\pmb{s}}} \tilde{\tau}_{\pmb{w}}^{\pmb{\nu}}.
\end{equation}
Notice that, unlike the noiseless classical shadows~\eqref{eq:noiseless-shadow-real}, the mitigated classical shadows $\tilde{\rho}^{\pmb{\nu}}_{\pmb{s}}$ do not factorise as tensor products over the qubits.
It is thus not obviously that transformation of the measurement outcomes to the sampling shades of the correlation function $c^{\pmb{\mu}}_{\pmb{v}}$ can be carried out efficiently.
Fortunately, this turns out still to be the case.

Recall that for a correlator $C^{\pmb{\mu}}_{\pmb{v}}$, the shade created by a classical shadow $\tilde{\rho}^{\pmb{\nu}}_{\pmb{s}}$  is defined by
$c^{\pmb{\mu},\pmb{\nu}}_{\pmb{v},\pmb{s}} =   \tr(\tilde{\rho}^{\pmb{\nu}}_{\pmb{s}} C^{\pmb{\mu}}_{\pmb{v}})$.
We then have
\begin{equation}
c^{\pmb{\mu},\pmb{\nu}}_{\pmb{v},\pmb{s}} = \frac{1}{2^n} \sum_{\pmb{w}}  (-1)^{\dprod{\pmb{w}}{\pmb{s}}} \frac{1}{g(\pmb{w})} \tr[\tau^{\pmb{\nu}}_{\pmb{w}} C^{\pmb{\mu}}_{\pmb{v}}].
\end{equation}
Interestingly, for the product correlator $C^{\pmb{\mu}}_{\pmb{v}}$ and for classical shadows~\eqref{eq:noiseless-shadow-fourier-explicit}, this exponential summation is reduced to a single term for the wavevector that matches the correlator pattern, $\pmb{w}=\pmb{v}$.
Indeed, inserting 
    $C^{\pmb{\mu}}_{\pmb{v}} = \otimes_{i=1}^{n} [\delta_{v_i,0} \openone + \delta_{v_i,1} \sigma^{\vec{\mu}_i}]$, 
we find the  formula for the mitigated shade claimed in the main text
\begin{equation}
c^{\pmb{\mu},\pmb{\nu}}_{\pmb{v},\pmb{s}} =  \frac{(-1)^{\dprod{\pmb{v}}{\pmb{s}}}}{g(\pmb{v})} \prod_{i=1}^{n}  [\delta_{v_i,0}  +  \delta_{v_i,1}\frac{1}{2} \tr(\xi^{\vec{\nu}_i} \sigma^{\vec{\mu}_i})].
    \label{eq:noisy-shade}
\end{equation}

\section{Sample complexity for protocols for $X$-twirled shadow tomography}

\subsection{Calibration and its complexity}

The calibration protocol given in the main text outputs calibration bitstring $\{\pmb{s}^{(k)}\}_{k=1}^{M_{\text{c}}}$. which can be used to estimate the Fourier component of the twirled noise $g(\pmb{w})$ by
\begin{equation}
    \hat{g}(\pmb{w}) = \frac{1}{M_c} \sum_{k=1}^{M_c}  (-1)^{\dprod{\pmb{w}}{\pmb{s}^{(k)}}}.
    \label{eq:gw-estimate}
\end{equation}
The calibration is in fact the same as that for mitigation of direct measurements of observables by $X$-twirling~\cite{Berg2022a}.
The sample complexity analysis below follows in a similar way.

From~\eqref{eq:gw-estimate}, observe that $\hat{g}(\pmb{w})$ is estimated as the mean value of a random variable taking values of $\{-1,1\}$.
The complexity in estimating $\hat{g}(\pmb{w})$ thus follows the general Hoeffding inequality~\cite{Hoeffding1963a}.
However, the estimation of observables from classical shadows requires the inverse of $g(\pmb{w})$. Unfortunately, for large $\abs{\pmb{w}}$, $g(\pmb{w})$ are generally small and we are facing with the estimation of the inverse of a small quantity. 

One starts by observing that~\cite[Lemma 1]{Berg2022a} 
\begin{equation}
    \abs{1/\hat{g} (\pmb{w}) - 1/g (\pmb{w})} \le \epsilon 
\end{equation}
holds if 
\begin{equation}
    \abs{\hat{g} (\pmb{w}) - g (\pmb{w})} \le \epsilon g (\pmb{w})/4.
\end{equation}

Then one can apply the Hoeffding inequality~\cite{Hoeffding1963a} to show that 
\begin{equation}
    \Pr \{ \abs{ \hat{g}(\pmb{w}) - g(\pmb{w})} \ge g (\pmb{w}) \epsilon/4 \} \le 2 \exp\{- M_c \epsilon^2 g^2 (\pmb{w}) / 32 )\}.
\end{equation}
Therefore in order to guarantee
\begin{equation}
    \Pr \{ \abs{1/\hat{g}(\pmb{w}) - 1/g(\pmb{w})} \ge \epsilon \} \le \delta
    \label{eq:gw-bound}
\end{equation}
one would requires
\begin{equation}
    M_{\text{c}} > -  32 \ln (\delta/ 2) /\epsilon^2 \times 1/g^{2} (\pmb{w}).
    \label{eq:calibration-complexity}
\end{equation}

One sees that the number of samples requires to estimate $1/g (\pmb{w})$ scales with $1/g^{2} (\pmb{w})$. To have an idea of how $g (\pmb{w})$ behaves, one can take the simple model of independent-flips~\eqref{eq:factorised-noise}. Further, one assumes that the observed outcome is obtained from the ideal bitstring outcome by flipping every bits with rate $\eta$. Direct calculation then show that $g (\pmb{w}) = (1- 2 \eta)^{\abs{w}}$. 
This illustrates that $g(\pmb{w})$ decays exponentially as the wavevector $\pmb{w}$ measured in the Hamming distance increases, thus harder to estimate.

\subsection{Shadow tomography and its complexity}
\label{sec:complexity}

The shadow tomography protocol in the main text outputs the sampled generalised outcomes $\{(\pmb{\nu}^{(l)},\pmb{s}^{(l)})\}_{l=1}^{M_{\text{st}}}$, which can be used to estimate correlation functions  in downstream information processing.
For a correlator $C^{\pmb{\mu}}_{\pmb{v}}$, the expectation value is estimated by 
\begin{equation}
    \hat{c}^{\pmb{\mu}}_{\pmb{v}} = \frac{1}{M_{\text{st}}} \sum_{l=1}^{M_{\text{st}}}
    \frac{(-1)^{\dprod{\pmb{v}}{\pmb{s}^{(l)}}}}{\hat{g}(\pmb{v})} \prod_{i=1}^{n}  [\delta_{v_i,0}  +  \delta_{v_i,1}\frac{1}{2} \tr(\xi^{\vec{\nu}_i^{(l)}} \sigma^{\vec{\mu}_i})]
\end{equation}

One might already expect that sample complexity for shadow tomography is significantly higher than the calibration process. 
Therefore in order to estimate the sample complexity for shadow tomography, we can assume that $g^{-1}(\pmb{v})$ can be estimated with high accuracy.

To estimate the sample complexity for shadow tomography, we observe that the shade $c^{\pmb{\mu},\pmb{\nu}}_{\pmb{v},\pmb{s}}$ is bounded between $- \kappa^{\abs{\pmb{v}}}/g(\pmb{v})$ and $+ \kappa^{\abs{\pmb{v}}}/g(\pmb{v})$, where $\kappa= 1/2 \max  \tr(\xi^{\vec{\nu}_i^{(l)}} \sigma^{\vec{\mu}_i})$. For the shadow tomography with Pauli measurements, one has $\kappa=3$. 
The accuracy of the estimated value $\hat{c}^{\pmb{\nu}}_{\pmb{s}}$ can then be bounded using the Hoeffding inequality~\cite{Hoeffding1963a}
\begin{equation}
    \operatorname{Pr}\{\abs{\hat{c}^{\pmb{\nu}}_{\pmb{s}} - c^{\pmb{\nu}}_{\pmb{s}}} \ge \epsilon \} \le 2 \exp \{ - M_{\text{st}}  [\epsilon g(\pmb{w})]^2 /(2\kappa^{2\abs{v}}) \}.
\end{equation}
Therefore in order to guarantee 
\begin{equation}
    \operatorname{Pr}\{\abs{\hat{c}^{\pmb{\nu}}_{\pmb{s}} - c^{\pmb{\nu}}_{\pmb{s}}} \ge \epsilon \} \le \delta,
\end{equation}
it is required that
\begin{equation}
    M_{\text{st}} > -2 \ln (\delta/2)/\epsilon^2 \times   \kappa^{2\abs{v}}/g^{2}(\pmb{w}).
    \label{eq:st-complexity}
\end{equation}
One also observes that the factor $\kappa^{2\abs{v}}$ contributes an exponential scaling of the required  number of samples for shadow tomography. 
This is known in shadow tomography~\cite{huang_predicting_2020}.
As regarding the effect of the readout noise, the required samples scales as $1/g^{2}(\pmb{w})$, similarly to the complexity of the calibration process.
Notice that we have used Hoeffding inequality to obtain a rather conservative scaling of the sample complexity. Tighter bounds for the sample complexity can be expected by better estimating the variance of the estimators by means of the so-called shadow norm as typically considered in shadow tomography, see, e.g.,  Ref.~\cite{huang_predicting_2020}.

\bibliography{stomography}

\end{document}